\documentclass{article}

\usepackage{arxiv}

\usepackage[utf8]{inputenc} % allow utf-8 input
\usepackage[T1]{fontenc}    % use 8-bit T1 fonts
\usepackage{hyperref}       % hyperlinks
\usepackage{url}            % simple URL typesetting
\usepackage{booktabs}       % professional-quality tables
\usepackage{amsfonts}       % blackboard math symbols
\usepackage{microtype}      % microtypography
\usepackage{lipsum}		% Can be removed after putting your text content
\usepackage{graphicx}
\usepackage{amsmath}
\usepackage{eucal}
\usepackage{siunitx}

\title{\emph{gnlse-python}: Open Source Software to Simulate Nonlinear Light Propagation In Optical Fibers}

\author{
    {Paweł Redman}\\
	Department of Optics and Photonics\\
	Wrocław University of Science and Technology\\
	\texttt{pawel.redman@gmail.com} \\
	\And
	{Magdalena Zatorska} \\
	Department of Optics and Photonics\\
	Wrocław University of Science and Technology\\
	\texttt{magdalena.zatorska@pwr.edu.pl} \\
	\And
	{Adam Pawłowski}\\
	Department of Optics and Photonics\\
	Wrocław University of Science and Technology\\
	\texttt{adam.97@vp.pl} \\
	\And
	{Daniel Szulc} \\
	Department of Optics and Photonics\\
	Wrocław University of Science and Technology\\
	\texttt{daniel.szulc@pwr.edu.pl} \\
	\And
	{Sylwia Majchrowska}\\
	Department of Optics and Photonics\\
	Wrocław University of Science and Technology\\
	\texttt{sylwia.majchrowska@pwr.edu.pl} \\
	\And
	{Karol Tarnowski} \\
	Department of Optics and Photonics\\
	Wrocław University of Science and Technology\\
	\texttt{karol.tarnowski@pwr.edu.pl} \\
}

\begin{document}
\maketitle

\begin{abstract}
The propagation of pulses in optical fibers is described by the generalized nonlinear Schrodinger equation (GNLSE), which takes into account the fiber losses, nonlinear effects, and higher-order chromatic dispersion.
The GNLSE is a partial differential equation, whose order depends on the accounted nonlinear and dispersion effects.
We present \emph{gnlse-python}, a nonlinear optics modeling toolbox
that contains a rich set of components and modules to solve the GNLSE using the split-step Fourier transform method (SSFM). The numerical solver is freely available, implemented in Python language, and includes a number of optical fiber analysis tools.
Code and data are available at \url{https://github.com/WUST-FOG/gnlse-python}.
%The project is under active development and we will keep this document updated.
\end{abstract}

% keywords can be removed

\section{Introduction}

Nonlinear optics is a branch of optics dealing with phenomena that are consequences of light-induced modifications to the optical properties of matter. Optical fibers are a very attractive medium for studying nonlinear phenomena. Due to the small size of the core and the geometry of the fibers, which allows light to propagate over very long distances, the waveguides can show nonlinearity even for the relatively low power of the initial pulses. Mathematical equations provide an excellent description of the nonlinear effects observed in this type of medium. Unfortunately, usually they are so complicated that they do not have an analytical solution, and the use of numerical integration techniques is unavoidable.

The numerical tools offer an excellent opportunity for deep inspection of light propagation in the computational investigation of nonlinear phenomena occurring in optical fibers. They play a crucial role in supporting various types of laboratory experiments. First of all, numerical tools allow reducing the cost of experimental work. Moreover, they give access to all information about the propagating pulse on the whole propagation distance, not only measured at the output. It is also possible to switch different terms of the equation being solved on and off (including and excluding some specific phenomena) to understand the underlying physics.

The basic mathematical model describing the shaping of the envelope of the pulse amplitude is a NonLinear Schrödinger Equation (NLSE), which depends on the time and distance of propagation. It takes into account effects such as the second order dispersion and phase self-modulation. It is successfully used in modeling the picosecond propagation of low-energy pulses propagating in optical networks. However, it is insufficient for femtosecond high energy pulses. To improve the existing NLSE model, the dispersion of higher orders and expressions describing successive nonlinear processes (e.g., Raman scattering) were added, yielding the GNLSE~\cite{bib:A19}.
Numerical modeling of the propagation of light pulses in various optical fibers based on GNLSE is used to describe complex phenomena such as supercontinuum generation. In practice, the equation facilitates the design of optical fibers, allowing for the generation of coherent radiation.

Considering the fact that the numerical solution of this complicated partial differential equation is quite time-consuming, it is necessary to develop efficient algorithms to solve it. High accuracy and relatively small inference time of the constructed methods is required. In the report, we introduce the open source Python toolbox for solving the GNLSE, based on the split-step Fourier transform method (SSFM), which takes the least computation time among all compared numerical schemes for GNLSE when the solution varies slowly with time~\cite{bib:AZS05}.

\section{Numerical model}
\label{sec:model}

In general, the GNLSE is an example of an equation that generally does not have analytical solutions. Moreover, the use of a numerical approach is necessary to solve the GNLSE and to describe the nonlinear processes occurring in optical fibers. In the time domain, the GNLSE takes the form

\begin{equation}
\begin{split}
    \frac{\partial A(z,T)}{\partial z} = & -\frac{\alpha}{2}A(z,T) - 
    \sum_{k\geq 2} \left(\frac{i^{k-1}}{k!}\beta_k
    \frac{\partial^k A(z,T)}{\partial T^k} \right) + \\
    & + i\gamma\left(1 + \frac{i}{\omega_0} \frac{\partial}{\partial T} \right)
    \left( A(z,T) \int_{-\infty}^{+\infty} R(T') |A(z,T-T')|^2 dT' \right),
\end{split} \label{eq:gnlse}
\end{equation}

where $\alpha$ is the attenuation constant, and $\beta_k$ are the higher order dispersion coefficients obtained by a Taylor series expansion of the propagation constant $\beta(\omega)$ around the center frequency $\omega_0$. The term on the second line describes the nonlinear effects --
time derivative in this term is responsible for self-steepening and optical wave breaking, whereas the convolution integral describes the delayed Raman response $R(T')$~\cite{bib:H07}. This form of the GNLSE is commonly employed for numerical simulations of pulse propagation in
a nonlinear medium such as an optical fiber.

Eq.~\ref{eq:gnlse} can also be expressed in the frequency domain by overlaying the Fourier transform as~\cite{bib:A19}
\begin{equation}
\begin{split}	
    \frac{\partial A(z,\omega)}{\partial z} = & -\frac{\alpha}{2}A(z,\omega) + \sum_{k\geq 2}\frac{i\beta_k}{k!}\omega^k A(z,\omega) + \\
    & + i\gamma\left(1 + \frac{\omega}{\omega_0} \right)
    \mathcal{F}\left\{ A(z,T) \int_{-\infty}^{+\infty} R(T') |A(z,T-T')|^2 dT'\right\}.
\end{split}
\label{eq:GNLSEF}
\end{equation}
The first and second terms on the right-hand side of Eq.~\ref{eq:gnlse} and Eq.~\ref{eq:GNLSEF} indicate the dispersion and nonlinear terms, respectively.

The SSFMs are a class of numerical methods that allow for the study of the behavior of a light pulse propagating in an optical fiber. To better understand the idea behind the algorithm, Eq.~\ref{eq:gnlse} can be written in the following form

\begin{equation}
\frac{\partial A}{\partial z} = \left( \hat{D} + \hat{N} \right)A,
\end{equation}

where $\hat{D}$ is the dispersion operator showing the influence of higher order dispersion and attenuation on the pulse propagation, while $\hat{N}$ is the nonlinear operator describing nonlinear phenomena occurring in the optical fiber. They are properly defined in both the time and frequency domains.

In general, both dispersion and nonlinearity act simultaneously along the entire length of the fiber. However, in the SSFM, we assume that they act independently at a short distance ($h$). In other words, the propagation along the segment $[z, z + h]$ can be modeled first by using the nonlinearity operator $\hat{N}$ (we assume that $\hat{D} = 0$), and then the dispersion operator $\hat{D}$. This leads to the approximation~\cite{bib:A19}

\begin{equation}	
A(z+h,T) \approx \exp(h\hat{D})\exp(h\hat{N})A(z,T).
\label{eq:ssfm}
\end{equation}

The dispersion operator is easier to calculate in the frequency domain to which we transfer using the Fourier transform. The return to the time domain is performed using the inverse Fourier transform. In the case of SSFM, the linear (dispersive) term is exactly accounted for, as it is completely integrable, whereas the nonlinear term is usually integrated using Runge-Kutta (RK) algorithms. A detailed description about the technique employed here can be found in~\cite{bib:H07, bib:DT10}.

\section{Toolbox deliverables}
\label{sec:implementation}

In our framework, we proposed an efficient solver, thanks to an adaptive step size implementation of
the fourth-order Runge-Kutta in the Interaction Picture method (RK4IP), adapted from~\cite{bib:H07, bib:DT10}. Numerical integration of a system of ordinary differential equations is done by \textit{solve\_ivp} from the \textit{scipy.integrate} package. Our Python code is partially based on MATLAB code published in~\cite{bib:DT10}, and available at~\url{http://scgbook.info/}. The toolbox prepares the integration using \textit{SciPy's} ODE solver, while the transitions between time and frequency domains are accomplished using the FFT and iFFT from \textit{pyfftw} library.

The solver module is divided into three parts:
\begin{itemize}
    \item specific model setup (\textit{gnlse.GNLSESetup}),
    \item the framework for preparing the split-step Fourier algorithm (\textit{gnlse.GNLSE}),
    \item and a class for managing solutions (\textit{gnlse.Solution}).
\end{itemize}

\subsection{Initial conditions: pulse envelopes}
\label{sec:envelopes}

To follow the evolution of the amplitude envelope in time and frequency, it is necessary to set appropriate boundary conditions. In this case, it is the known initial pulse shape described by the functions $A(0,T)$ or $A(0,\omega)$. They represent the complex envelope amplitude of the pulse at the fiber input.

\begin{figure}[!h]
	\centering
	\includegraphics[width=.95\textwidth]{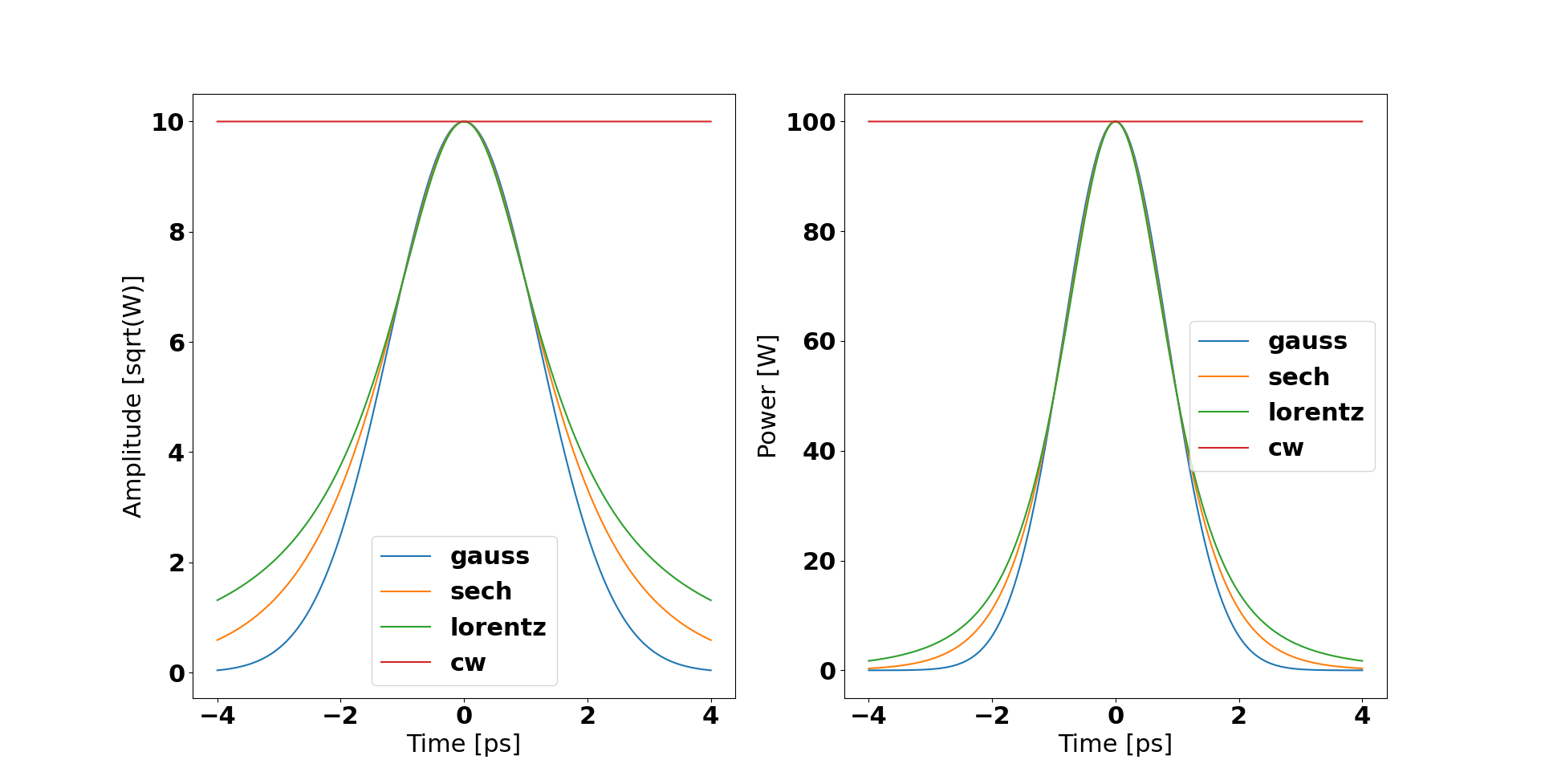}
	\caption{Example input pulse envelops in time domain.}
	\label{fig:envelopes}
\end{figure}

In physics, the envelope of an oscillating signal is a smooth curve outlining its extremes. The given envelope is a function of time, describing the amplitude of the input pulse calculated basing on its specific properties, i.e., pulse duration full-width half-maximum and peak power (or only average power for a continuous wave). Fig.~\ref{fig:envelopes} illustrates the envelopes of four types of optical signals implemented in the \textit{envelopes} module. We allow the user to choose between Gaussian, Lorentzian, hyperbolic secant-shaped envelopes and a continuous wave with optional noise.

\subsection{Dispersion operator}
\label{sec:dispersion}

Dispersion is a phenomenon that manifests itself as the dependence of the phase velocity of a wave on other quantities, such as frequency or polarization. One of the most important parameters characterizing optical fibers is the chromatic dispersion, which expresses the dependence of the signal propagation velocity on the wavelength. Chromatic dispersion plays a huge role in the propagation process of a pulse in an optical fiber.
Different frequency components have different propagation constants $\beta$, which can be expressed by the equation \cite{bib:A19}
\begin{equation} 
    \beta = \frac{n\lambda}{2\pi}.
\end{equation}
Since the frequencies present in time-limited pulses are concentrated around the central frequency $\omega_0$, we can use the expansion of $\beta(\omega)$ into the Taylor series
\begin{equation} 
\beta(\omega) = n(\omega)\frac{\omega}{\mathrm{c}} = \beta_0 + \beta_1(\omega-\omega_0) + \frac{1}{2}\beta_2(\omega-\omega_0)^2 + ... ,
\end{equation}
where
\begin{equation} 
\beta_m = \left(\frac{\mathrm{d}^m\beta}{\mathrm{d}\omega^m}\right)_{\omega = \omega_0}.
\end{equation}

For such representation, the dispersion operator in the frequency domain is defined as
\begin{equation} 
\hat{D} =  \sum_{k\geq 2}\frac{i\beta_k}{k!}\left(\omega-\omega_0\right)^k -\frac{\alpha}{2},
\label{eq:summd}
\end{equation}
where the second term of the right-hand side is responsible for loss.

The $\beta_1$ and $\beta_2$ parameters are defined as
\begin{equation} 
    \beta_1 = \frac{1}{v_g} = \frac{n_g}{\mathrm{c}} = \frac{1}{\mathrm{c}}\left(n + \omega\frac{\mathrm{d}n}{\mathrm{d}\omega}\right)\,\,\,\,\, \mathrm{\left[\frac{ps}{km}\right]},
\end{equation}
\begin{equation} 
    \beta_2 = \frac{1}{\mathrm{c}}\left(2\frac{\mathrm{d}n}{\mathrm{d}\omega}+\omega\frac{\mathrm{d}^2n}{\mathrm{d}\omega^2}\right)\,\,\,\,\, \mathrm{\left[\frac{ps^2}{km}\right]},
\end{equation}
where $v_g$ is the group velocity at which the pulse propagates, and $n_g$ is the group refractive index. Thus, $\beta_1$ is the reciprocal of the group velocity, and $\beta_2$ quantifies the broadening of the pulse due to the fact that not all of its components propagate at the same velocity. This phenomenon is called group velocity dispersion (GVD). Higher orders of dispersion are also defined; however, for pulses with narrow spectral bands, they can be omitted. In most cases, it is enough to take into account the $\beta_3$ parameter, which is responsible for the phenomenon of third-order dispersion. Taking higher order dispersion into account is usually necessary in the case of extreme spectral broadening, such as in the supercontinuum generation process.
        
The linear operator in frequency domain can be also expressed by~\cite{bib:A19}
\begin{equation}
\hat{D} = i\left\{
	\beta(\omega) - \left[
		\beta(\omega_0) + \beta_1(\omega_0)(\omega-\omega_0)
		\right]
	\right\} - \frac{\alpha(\omega)}{2},
	\label{eg:disp}
\end{equation}
where $\omega_0$ is both the center frequency of the defined grid and the central frequency $\omega_c$ corresponding to the wavelength of the central component of the spectrum of the introduced pulse.
		
The first term of the expression above corresponds to the Taylor series expansion (starting from the second term) of the $\beta$ propagation constant
		
\begin{equation}    	
	\beta(\omega) - \left[
		\beta(\omega_0) + \beta_1(\omega_0)(\omega-\omega_0)
		\right] = 
		\sum_{k\geq 2}\frac{(\omega -\omega_0)^k}{k!}\left(\frac{\mathrm{d}^k\beta}{\mathrm{d}\omega^k}\right)_{\omega = \omega_0}.
\end{equation}

The dispersion operator defined by Eq.~\ref{eg:disp} is extremely useful when we know the exact values of the propagation constant $\beta$, not just its successive derivatives. In the case of a known dependence of the effective refractive index on the frequency, it is enough to make appropriate calculations using the formula

\begin{equation} 
\beta(\omega) = n_\mathrm{eff}(\omega)\frac{\omega}{\mathrm{c}}.
\end{equation}
The next step is to calculate the value of $\beta$ and its first derivative at $\omega_0 $.

\begin{figure}[!h]
	\centering
	\includegraphics[width=\textwidth]{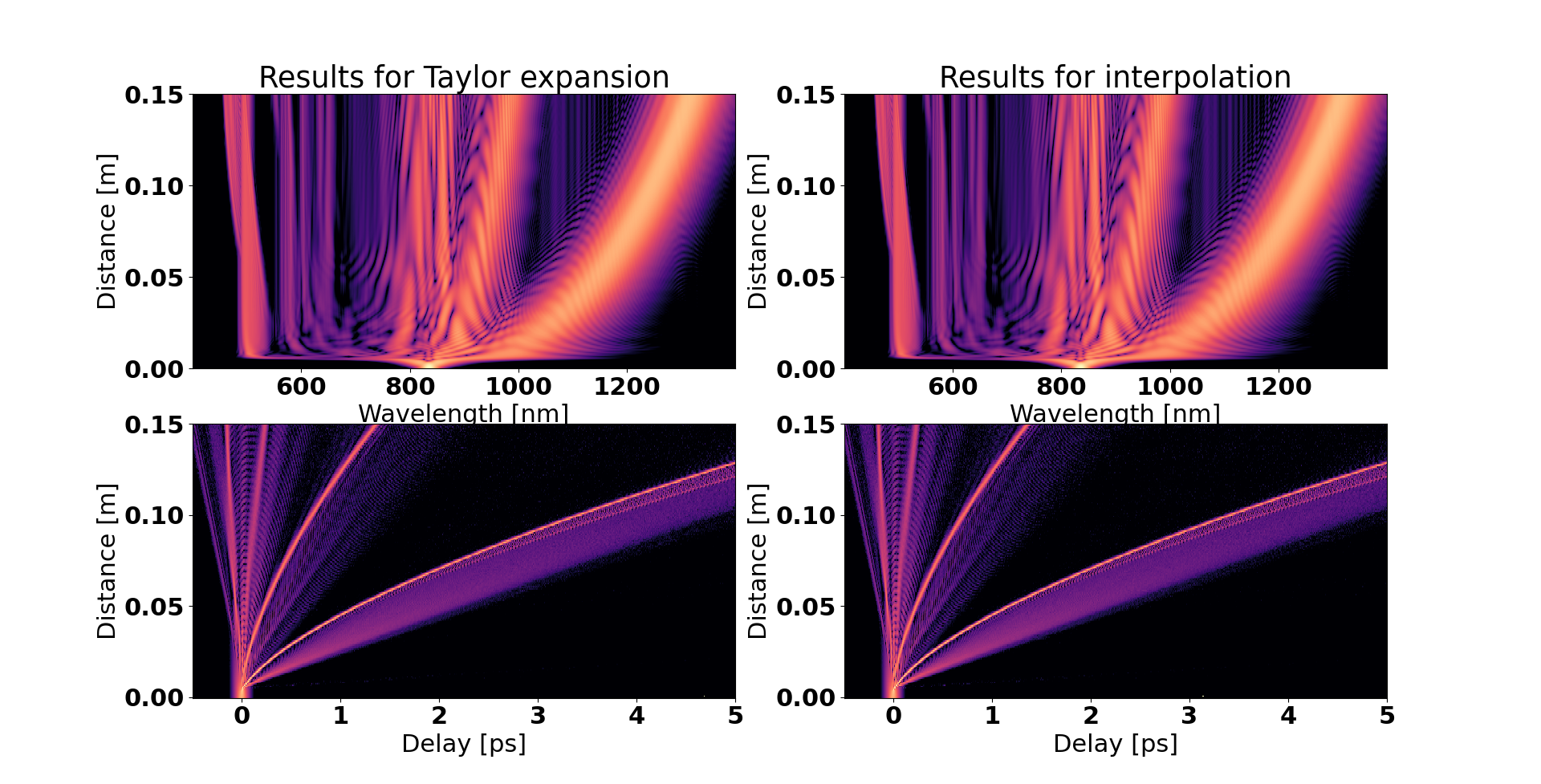}
	\caption{Example of supercontinuum generation using different dispersion operators. In our example we used provided $\beta_p$ coefficients up to the 10th derivative.}
	\label{fig:disp}
\end{figure}

Both forms of the dispersion operator (namely, Eq.~\ref{eq:summd}, and Eq.~\ref{eg:disp}) have been implemented in the described toolbox. In case of taking into account the full dependence of propagation constants on frequency, we use an extrapolation method for the provided effective refractive indices (and calculate propagation constants from them) from an external file. For this purposes the \textit{interp1d} function from \textit{SciPy} library was used. A demonstration of the influence of accounted input parameters is shown in Fig.~\ref{fig:disp} for the supercontinuum generation example.

\subsection{Nonlinear phenomena: Raman responses}
\label{sec:raman}

The last term in Eq.~\ref{eq:gnlse} describes nonlinear phenomena such as: phase self-modulation, self-steepening, shock wave formation, and Raman scattering. The $R(t)$ function, which determines the last of the listed effects, usually is modelled as~\cite{BW89}
\begin{equation} 
R(t) = (1 - f_R)\delta(t) + f_Rh_R(t),
\label{eq:rammanowska}
\end{equation}  
where
\begin{equation} 
h_R(t) = \Theta(t)\frac{\tau_1^2 + \tau_2^2}{\tau_1\tau_2^2}\exp\left(-\frac{t}{\tau_2}\right)\sin\left(\frac{t}{\tau_1}\right). \label{eq:hR}
\end{equation}  
$\Theta(t)$ is a Heaviside step function, while $f_R$, $\tau_1$ and $\tau_2$ depend on the type of material. For silica glass, they equal $f_R = 18\%$, $\tau_1 = \SI{12.2}{\femto\second}$ and $\tau_2 = \SI{32}{\femto\second}$.

\begin{figure}[!h]
	\centering
	\includegraphics[width=\textwidth]{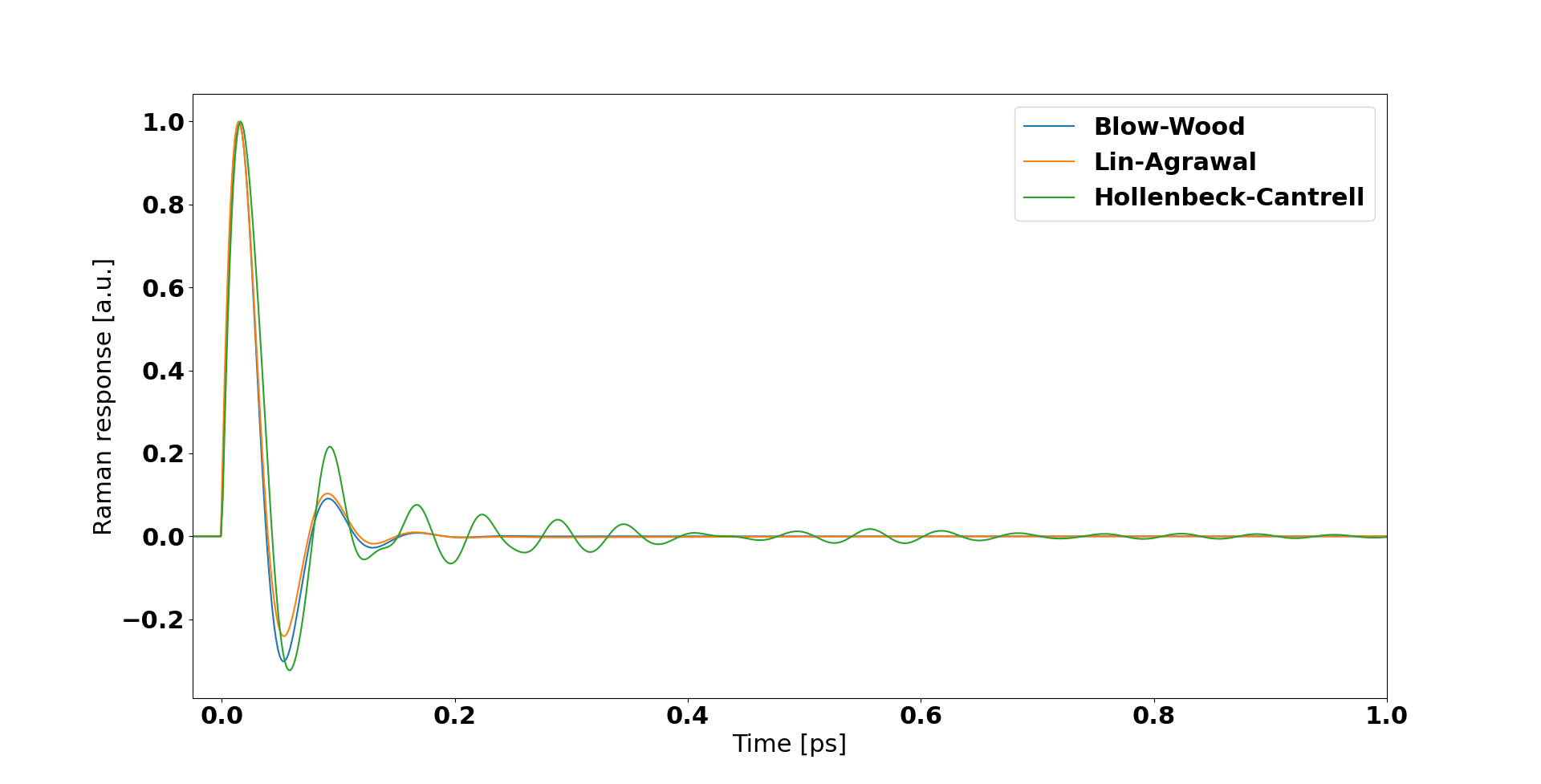}
	\caption{Raman scattering functions for silica optical fibers for three models in time domain.}
	\label{fig:raman}
\end{figure}

The response is calculated based on the chosen Raman model. There are three available Raman response models for silica optical fibers implemented: 
\begin{itemize}
    \item based on K. J. Blow and D. Wood model~\cite{BW89}, presented in Eq.~\ref{eq:rammanowska},
    \item and based on Q. Lin and Govind P. Agrawal model~\cite{LA06} as a more accurate response function,
    \item based on Dawn Hollenbeck and Cyrus D. Cantrell model~\cite{HC02}, which presents the multi-vibrational-mode function,
\end{itemize}
Their visualisations in the time domain are presented in Fig.~\ref{fig:raman}.

\newpage
\subsection{Nonlinear phenomena: The nonlinear coefficient}
\label{sec:nonlinear}

The time derivative term inside GNLSE models the dispersion of
the nonlinearity. This is usually associated with effects such
as self-steepening and optical shock formation, characterized by
a timescale $\tau_0 = 1/\omega_0$. In the context of fibre
propagation, an additional dispersion of the nonlinearity arises
due to the frequency dependence of the effective mode area. The last effect can be accounted in $\tau_0$ coefficient in an approximate manner~\cite{BW89, K05}.

A better -- still approximate -- approach to include the dispersion of the effective mode area is to
describe it directly in the frequency domain~\cite{L07}.
In this case, we can derive a GNLSE for the pulse evolution using
$\overline{\gamma}(\omega)$ defined as
\begin{equation}
\begin{split}
   \overline{\gamma}(\omega) = &\,\frac{n_2n_{\mathrm{eff}}(\omega_0)\omega_0}{\mathrm{c}n_\mathrm{eff}(\omega)\sqrt{A_{\mathrm{eff}}(\omega)A_{\mathrm{eff}}(\omega_0)}}.
\end{split}
\end{equation} 

This approach is more rigorous than the approximation of ($\gamma = \gamma(\omega_0)$) and requires the definition of a pseudo-envelope $C(z, \omega)$ as
\begin{equation} 
   C(z, \omega) = \left(\frac{A_{eff}(\omega_0 )}{A_{eff}(\omega )}\right)^{1/4} A(z, \omega).
   \label{eq:scale}
\end{equation}

\begin{figure}[!h]
	\centering
	\includegraphics[width=\textwidth]{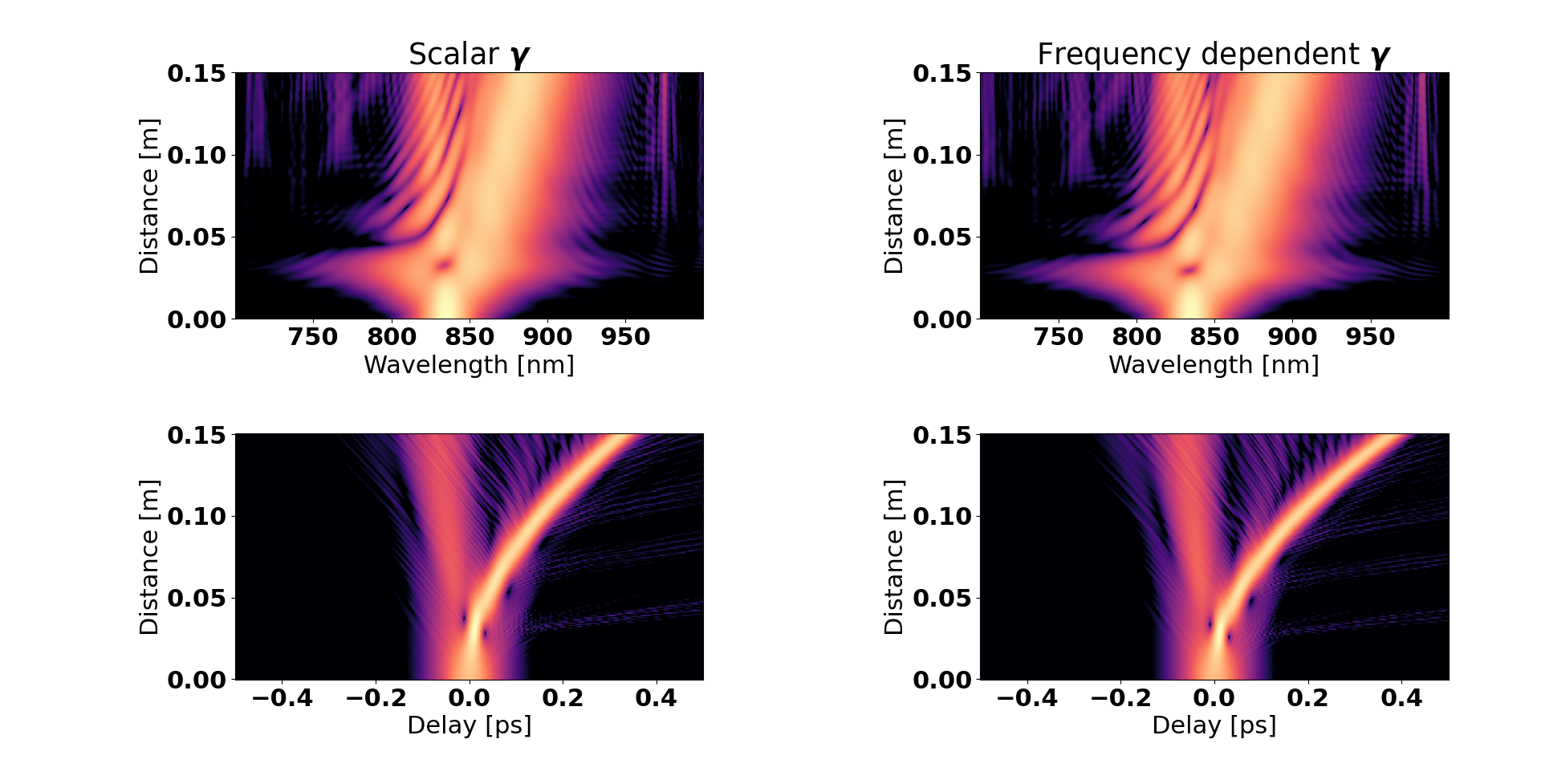}%{Figure_4.png}
	\caption{Example of soliton fission using different nonlinear coefficient values. In both cases the dispersion operators given by Taylor expansion of $\beta$ were used.}
	\label{fig:dispop}
\end{figure}

Both forms of the equation (namely, Eq.~\ref{eq:gnlse}, and the form introducing scaling Eq.~\ref{eq:scale}) have been implemented in the described toolbox. In the case of taking into account the full dependence of the nonlinear coefficient on frequency, we used an extrapolation method for the provided effective mode areas (and effective refractive indices) from an external file using \textit{interp1d} function from \textit{SciPy} library. Investigation of the influence of accounted input parameters is shown in Fig.~\ref{fig:dispop} for soliton fission example.

\newpage
\subsection{Additional tools}
\label{sec:tools}

We provide some additional tools and functions inside \textit{gnlse-python} package. As \textsc{MATLAB} is the most popular environment for nonlinear computations in the scientific community, we provided the module \textit{import\_export} that can handle \textit{\*.mat} files. Also the gnlse.Solution class has two methods to load and save the results to an external file of this type. For these purposes we used the \textit{hdf5storage} package.

In turn, the module \textit{visualization} is responsible for the visualization of the results. We chose the popular \textit{matplotlib} library for the task. We decided to show results of calculations as two-sided plots presenting a solution both in both the time and spectral domains. Default ranges for the x-axes correspond to the case of supercontinum generation as in work~\cite{bib:DT10}.

The \textit{common} module is the last stand-alone script, which in the future will keep the physical constants used in the calculations. Currently, we have placed there the value of speed of light in vacuum in nm/ps.

\section{Numerical illustrations}
\label{sec:examples}

We investigated the toolbox performance on some common examples presented in~\cite{bib:A19, bib:DT10}. We observe qualitative compliance with a relatively small calculation time.

\subsection{Propagation of higher-order soliton}

An electromagnetic wave scattered in all directions has different spectral components. The dominant band has the same frequency as the incident wave -- it is the Rayleigh band (elastic scattering). On the other hand, the components with frequencies shifted by the characteristic frequencies of molecules are called Raman scattering bands (inelastic scattering). These are the bands of reduced and increased frequencies, the Stokes and anti-Stokes bands, respectively. Anti-Stokes scattering occurs less frequently due to the lower occupation of states with higher oscillatory energy~\cite{bib:A19}.

Raman scattering can be both spontaneous or stimulated. The latter is dominant in the case of propagation of short pulses in the optical fiber. In optical fiber transmission, the stimulated Raman scattering can induce optical Raman gain. In addition, due to the forced Raman scattering, the spectrum shifts towards longer wavelengths, which is caused by the amplification of the Stokes wave with frequencies inside the spectrum. It is the so-called self-frequency shift phenomenon.

\begin{figure}[!h]
	\centering
	\includegraphics[width=.9\textwidth]{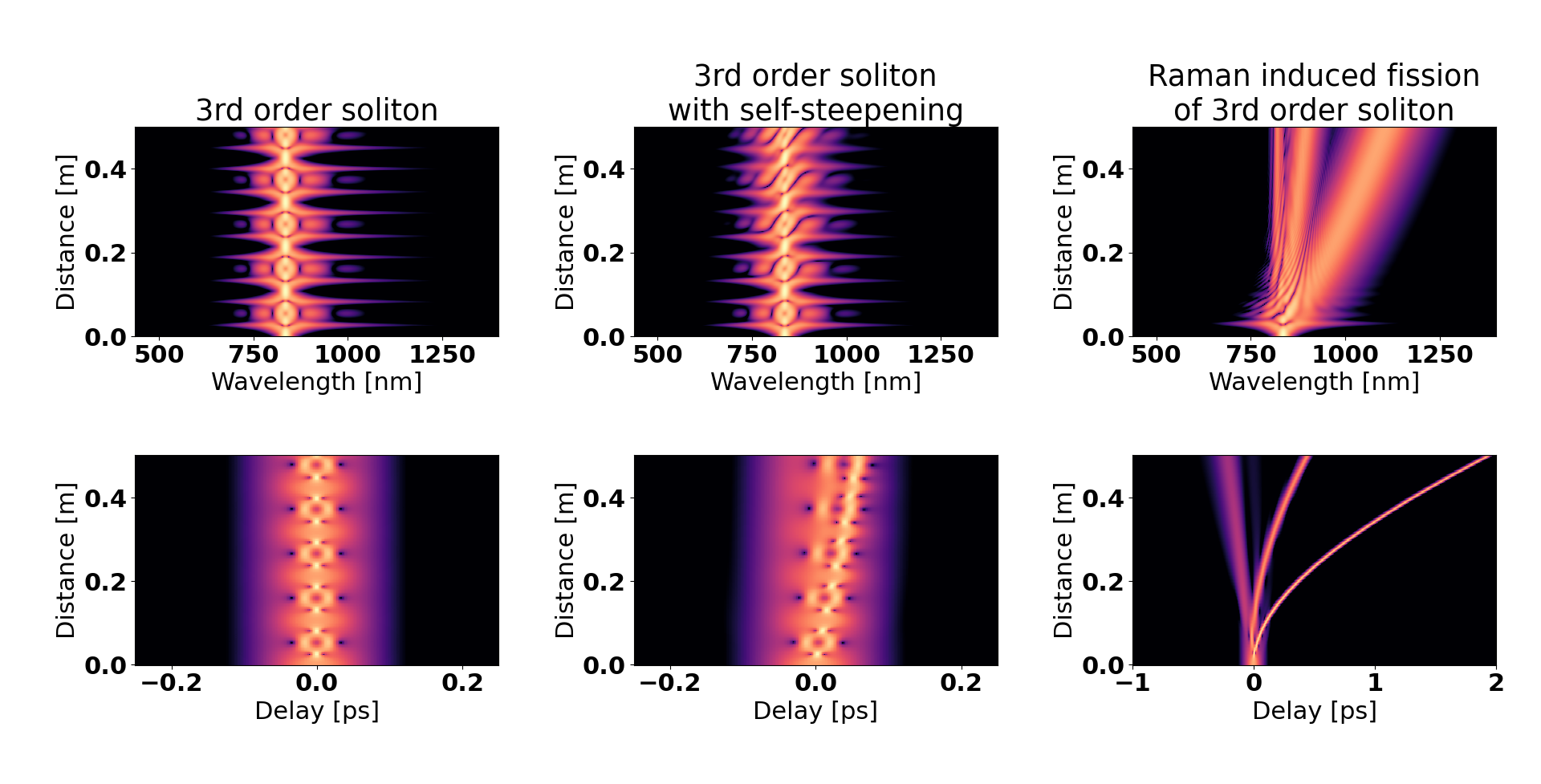}
	\caption{Evolution of the spectral and temporal characteristics of the higher-order N = 3 soliton in three cases.}
	\label{fig:soliton}
\end{figure}

The self-frequency shift phenomenon can be observed in the case of solitons -- optical pulses propagating in a~dispersive medium without changing their temporal and spectral intensity profiles. A higher-order soliton is a soliton with a higher energy than that of a fundamental soliton by a factor that is the square of an integer (soliton order). The temporal shape of such soliton varies periodically during propagation. The soliton can be also affected by the nonlinear increase of the pulse steepness (self-steepening phenomenon). Fig.~\ref{fig:soliton} shows the evolution of the spectral and temporal characteristics of a higher-order $N = 3$ soliton in three cases:
\begin{itemize}
    \item propagation without self-steepening and Raman response,
    \item soliton fission with self-steepening, but no Raman response accounted,
    \item soliton fission with self-steepening, and Raman response accounted,
\end{itemize}

\subsection{Supercontinuum generation in anomalous dispersion regime}

To a large extent, the interaction of an electromagnetic wave with the system of atoms in a medium depends on the optical properties of matter and the frequency of the incident light. In other words, quantities such as permittivity, refractive index, and absorption are functions of the wave frequency. These dependencies are called the dispersions, which in a broader sense also means the splitting of light into monochromatic components. It influences the phase matching when interacting with several waves~\cite{bib:A19}.

If we take a closer look at the light pulse, which cannot be treated as a monochromatic wave, it can be seen that the dispersion additionally changes its parameters. Due to the different velocities of the group components of the pulse frequency spectrum, the pulse's time profile is distorted. We can distinguish between two dispersion regimes: normal and anomalous. In the case of anomalous dispersion, the group refractive index increases with increasing wavelength -- the components with lower frequencies have a lower group velocity. 

\begin{figure}[!h]
	\centering
	\includegraphics[width=\textwidth]{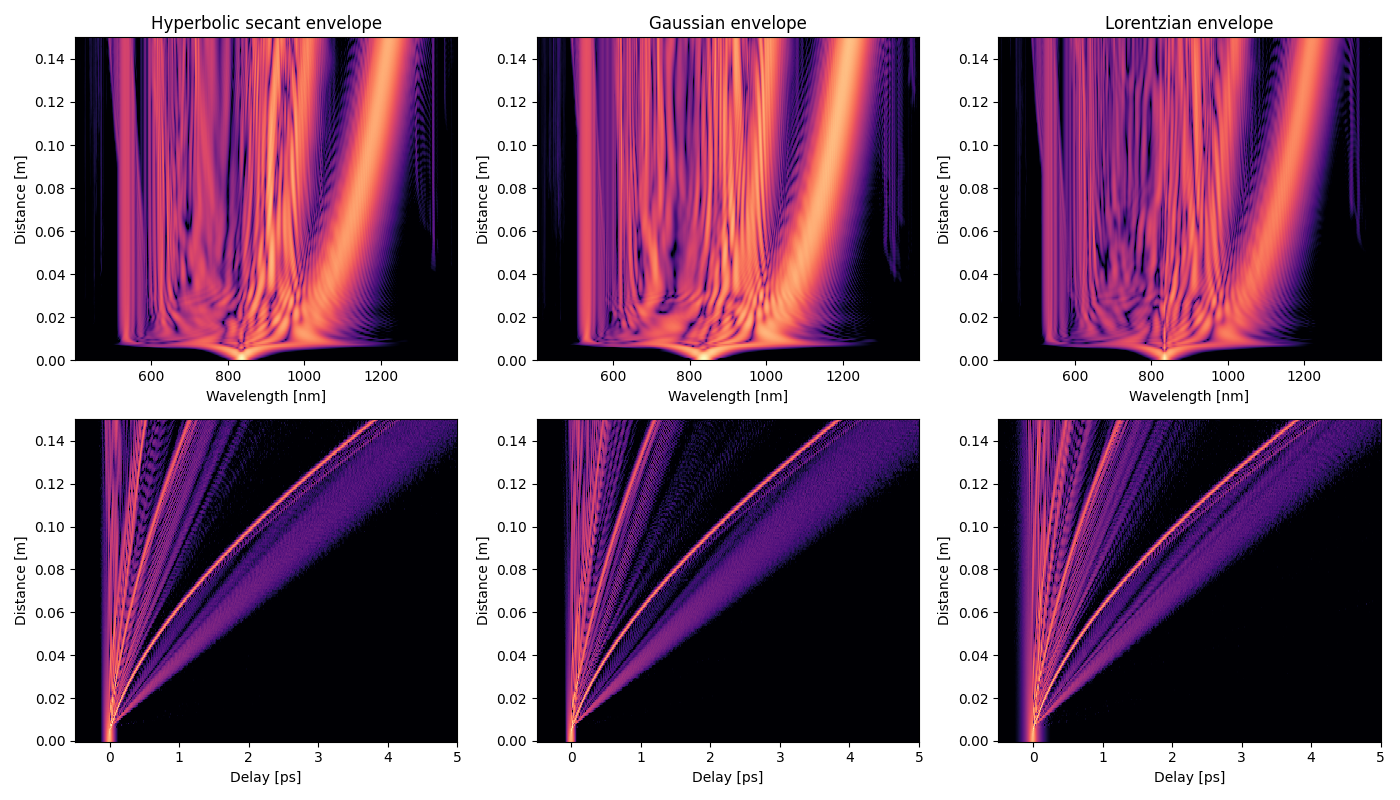}
	\caption{Example of supercontinuum generation in anomalous dispersion regime.}
	\label{fig:supercontinuum}
\end{figure}

Dispersion is rather undesirable in telecommunications systems, and the compensation of the anomalous dispersion can be achieved by optical nonlinearity, delaying the wave components at the beginning of the pulse and accelerating those at the end of~\cite{bib:A19}. In this way, we can try to compensate for the elongation of the pulse. Nonlinearity also causes many other useful effects that have found their practical application. One of them is the nonlinear spectral broadening process, which we call supercontinuum generation. Fig.~\ref{fig:supercontinuum} shows an example of supercontinuum generation in the anomalous dispersion regime at a central wavelength of 835 nm in a 15 centimeter long fiber using three different shapes of input pulse envelopes.

\newpage
\subsection{Dispersive wave generation in anomalous dispersion regime -- different Raman responses}

\begin{figure}[!h]
	\centering
	\includegraphics[width=.9\textwidth]{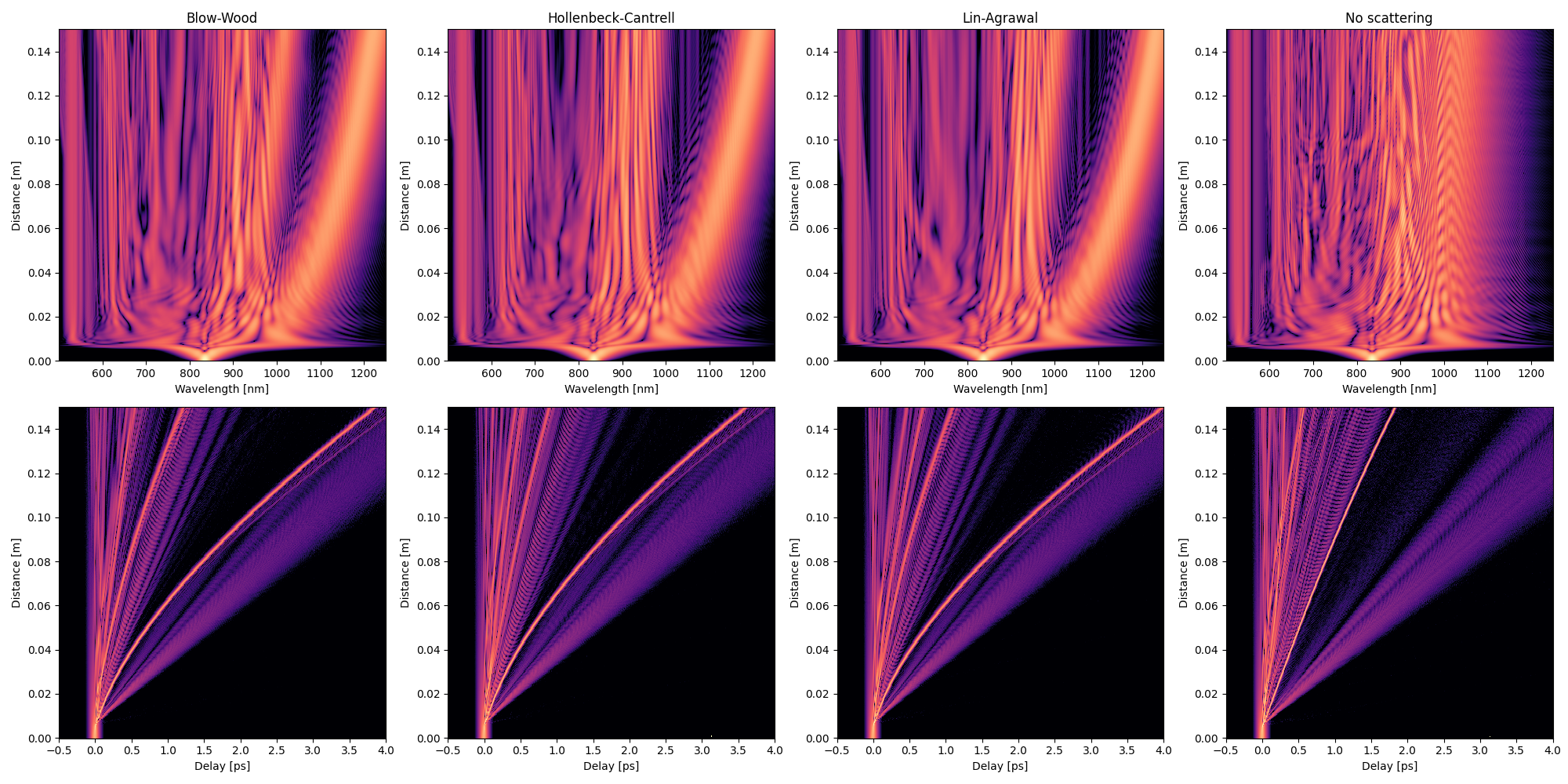}
	\caption{Example of dispersive wave generation in anomalous dispersion regime.}
	\label{fig:dispersive}
\end{figure}

The optical pulse propagating in a fiber with anomalous chromatic dispersion forms a soliton pulse accompanied with a~temporally spreading background -- the dispersive wave. The background radiation is generated in a high frequency band. The dispersive wave is spread by chromatic dispersion, which is not compensated by the fiber nonlinearity (due to low power). Fig.~\ref{fig:dispersive} shows an example of dispersive wave generation in anomalous dispersion regime at a central wavelength of 835 nm in a 15 centimeter long photonic crystal fiber using three different models of the Raman response.

\newpage
\section{Summary}

In this report, we present the open source \textit{gnlse-python} package written in the Python language. It is a set of Phyton scripts for solving the Generalized Nonlinear Schrodringer Equation. It is one of the WUST-FOG students projects developed by \href{http://www.fog.pwr.edu.pl}{Fiber Optics Group, WUST}.

The complete and up-to-date documentation of the project is available at \url{https://gnlse.readthedocs.io}.

\section*{Authors' contributions statement}

PR, MZ, AP, DS, SM, and KT contributed to the final code. PR is the solution architect, who proposed the structure of gnlse integration module based on the existing MATLAB code. MZ prepared a module implementing the initial pulses' temporal shapes, and a module for importing and exporting data. AP prepared an implementation of the Raman responses models and the dispersion operator interpolation components. DS took responsibility for implementing the visualization of the results. SM implemented the nonlinear module for accounting the dispersion of effective mode area. All authors were involved in the preparation of example visualization of physical phenomena. KT and SM supervised the project. SM was responsible for integrating the existing code with all separate modules and  prepared the technical document, while KT watched over the correct implementation from the theoretical point of view.

%\bibliographystyle{unsrt}
%\bibliography{references}

\end{document}